\documentclass[fleqn,10pt]{wlscirep}
\title{0.54 $\mu$m resolution two-photon interference with dispersion cancellation 
for quantum optical coherence tomography}

\author[1,2,3]{Masayuki Okano}
\author[4]{Hwan Hong Lim}
\author[1,2,3]{Ryo Okamoto}
\author[5]{Norihiko Nishizawa}
\author[4]{Sunao Kurimura}
\author[1,2,3,*]{Shigeki Takeuchi}
\affil[1]{Department of Electronic Science and Engineering, Kyoto University, Kyoto daigaku-katsura, Nishikyo-ku, Kyoto, Kyoto, Japan}
\affil[2]{Research Institute for Electronic Science, Hokkaido University, Sapporo, Hokkaido, Japan}
\affil[3]{The Institute of Scientific and Industrial Research, Osaka University, Mihogaoka 8-1, Ibaraki, Osaka, Japan}
\affil[4]{National Institute for Materials Science, 1-1 Namiki, Tsukuba, Ibaraki, Japan}
\affil[5]{Department of Quantum Engineering, Nagoya University, Furo-cho, Chikusa-ku, Nagoya, Aichi, Japan}

\affil[*]{takeuchi@kuee.kyoto-u.ac.jp}


\begin{abstract}
Quantum information technologies harness 
the intrinsic nature of quantum theory 
to beat the limitations of the classical methods 
for information processing and communication. 
Recently, the application of quantum features to metrology
has attracted much attention.
Quantum optical coherence tomography (QOCT),
which utilizes two-photon interference
between entangled photon pairs, 
is a promising approach to overcome the problem 
with optical coherence tomography (OCT): 
As the resolution of OCT becomes higher, 
degradation of the resolution due to dispersion within the medium
becomes more critical. 
Here we report on the realization of 0.54 $\mu$m resolution
two-photon interference, 
which surpasses the current record resolution 
0.75 $\mu$m of low-coherence interference for OCT. 
In addition, the resolution for QOCT showed almost no change 
against the dispersion of a 1 mm thickness of water 
inserted in the optical path,
whereas the resolution for OCT dramatically degrades.
For this experiment, 
a highly-efficient chirped quasi-phase-matched
lithium tantalate device 
was developed using a novel $`$nano-electrode-poling$'$ technique. 
The results presented here represent a breakthrough 
for the realization of quantum protocols, 
including QOCT, quantum clock synchronization, and more.
Our work will open up possibilities for medical and biological applications.
\end{abstract}
\begin{document}

\flushbottom
\maketitle
%
%
\thispagestyle{empty}

\section*{Introduction}

One of the most distinct feature of quantum physics is quantum entanglement. 
Entanglement attracted attention first in the test of nonlocality of quantum mechanics
\cite{{Einstein1935},{Bell1964},{Aspect1982}},
and started to be considered as an essential resource for quantum information protocols
\cite{Kimble2008},
including quantum key distribution \cite{{Ekert1991},{Gisin2007}},
quantum teleportation \cite{{Bennett1993},{Bouwmeester1997}},
and quantum computation \cite{{Knill2001},{Ladd2010},{Okamoto2010}}.
Recently, the application of quantum entanglement for metrology and sensing
is attracting attention \cite{{Giovannetti2004},{Nagata2007}}. 
One recent example is an entanglement-enhanced microscope,
where a photon-number entangled state is used as probe light
to enhance the sensitivity \cite{Ono2013}.
Here we focus on another example:
an application of the frequency entangled state of photons
for optical coherence tomography (OCT) \cite{{Huang1991},{Brezinski2006}}.

\begin{figure}[ht]
\begin{center}
\includegraphics[width=8.7cm]{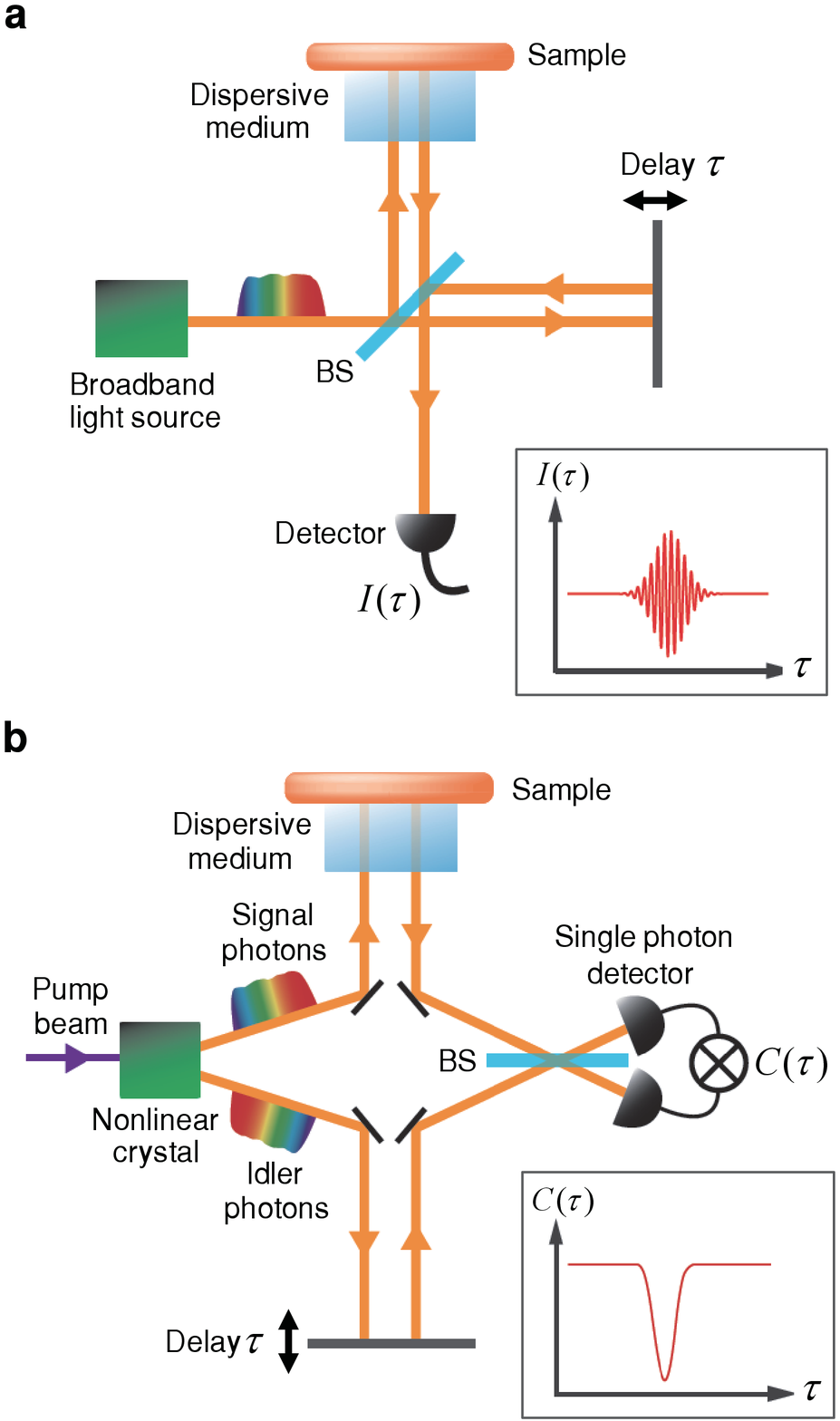}
\caption{OCT and QOCT schemes. 
(a), Schematic diagram of OCT. 
$I(\tau)$ is the interfered light intensity measured at a detector
with varying delay $\tau$ (inset).
BS is a beam splitter. 
(b), Schematic diagram of QOCT.
$C(\tau)$ is the coincidence count rate
counted at two single photon detectors
with varying delay $\tau$ (inset).}
\label{Fig1}
\end{center}
\end{figure}

OCT based on low-coherence interference (LCI) \cite{Max1999}
has been widely used in various fields, including medical applications 
such as imaging of the retina and cardiovascular system \cite{{Huang1991},{Brezinski2006}}. 
Figure \ref{Fig1}a shows a schematic diagram of OCT. 
Broadband light from a source is divided at a beam splitter (BS). 
One beam is incident on a sample after passing through a dispersive medium, 
while the other beam is reflected from a mirror with a temporal delay $\tau$. 
The OCT interference fringe $I(\tau)$ is obtained 
by the detection of the interfered light intensity with varying delay $\tau$ 
(Figure \ref{Fig1}a inset). 
When the bandwidth of the source is made broader to achieve higher resolution,
the resolution, far from being improved, degrades due to dispersion in the medium.
This constitutes a severe problem in OCT \cite{Hitzenberger1999}. 
Although dispersion effect can be compensated
by inserting a $`$phantom$'$,
a medium with the same dispersion, in the reference path \cite{Drexler2001},
it requires \textit{a priori} knowledge of both 
the structure and the frequency-dependent refractive index of the target object.
Furthermore, as the resolution becomes higher,
a slight difference between the target object and the $`$phantom$'$
becomes a crucial problem.

As an alternative method, 
quantum optical coherence tomography (QOCT) 
\cite{{Abouraddy2002},{Nasr2003},{Nasr2008},{Nasr2008-2},{Nasr2009}}
based on the two-photon interference (TPI) of 
frequency-entangled photon pairs \cite{Hong1987} 
has been proposed \cite{Abouraddy2002} (Figure \ref{Fig1}b). 
Broadband entangled photon pairs can be generated
via a spontaneous parametric down-conversion process from a nonlinear crystal.
Signal photons reflected at a sample through a dispersive medium 
and idler photons with a temporal delay $\tau$, 
interfere quantum mechanically at a BS, 
and coincidence detection events are counted at two single photon detectors. 
The QOCT interference dip $C(\tau)$, 
which is so called Hong-Ou-Mandel (HOM) dip \cite{Hong1987}, 
is obtained by the coincidence count rate with varying delay $\tau$ (Figure \ref{Fig1}b inset). 
Due to the frequency correlation of entangled photon pairs,
the resolution of QOCT (the width of the HOM dip)
does not change even with group velocity dispersion (GVD) in the medium
\cite{{Steinberg1992},{Steinberg1992-2}}.
This $`$dispersion cancellation$'$ of TPI was first demonstrated 
with 19 $\mu$m resolution \cite{Nasr2003}
and very recently with 3 $\mu$m resolution, 
where the GVD effect becomes significant \cite{Okano2013}. 

In this work, 
we report 0.54 $\mu$m resolution TPI with dispersion cancellation
for ultra-high resolution QOCT. 
For highly efficient generation of 
ultra-broadband (166 THz, $\lambda$ = 660-1040 nm) frequency-entangled photon pairs,
we developed a 1st-order chirped quasi-phase-matched (QPM) \cite{Harris2007} 
lithium tantalate \cite{Yu2004} device
using the nanofabrication technology for fine electrode patterns. 
The device was pumped using a narrowband ($\sim$ 100 kHz) pump laser
to ensure the dispersion cancellation of TPI \cite{Okano2013}.
We also constructed stable interferometer setup with
hybrid ultra-broadband detection systems (HUBDeS) operated at room temperature. 
In addition, 
dispersion cancellation in TPI was demonstrated against a 1 mm thickness of water
inserted in the optical path.
Almost no degradation in resolution was observed for TPI
(from 0.54 $\mu$m to 0.56 $\mu$m),
which is in contrast to the significant degradation in LCI resolution
(from 1.5 $\mu$m to 7.8 $\mu$m). 
The 0.54 $\mu$m resolution in air, 
which corresponds to the resolution of 0.40 $\mu$m in water, is the best 
among the previously achieved LCI resolution 0.75 $\mu$m \cite{Povazay2002} for OCT 
and also TPI resolutions for QOCT, including that 
(0.85 $\mu$m \cite{Nasr2008-2})
where dispersion cancellation was not verified. 

\section*{Results}

\begin{figure}[ht]
\begin{center}
\includegraphics[width=17.8cm]{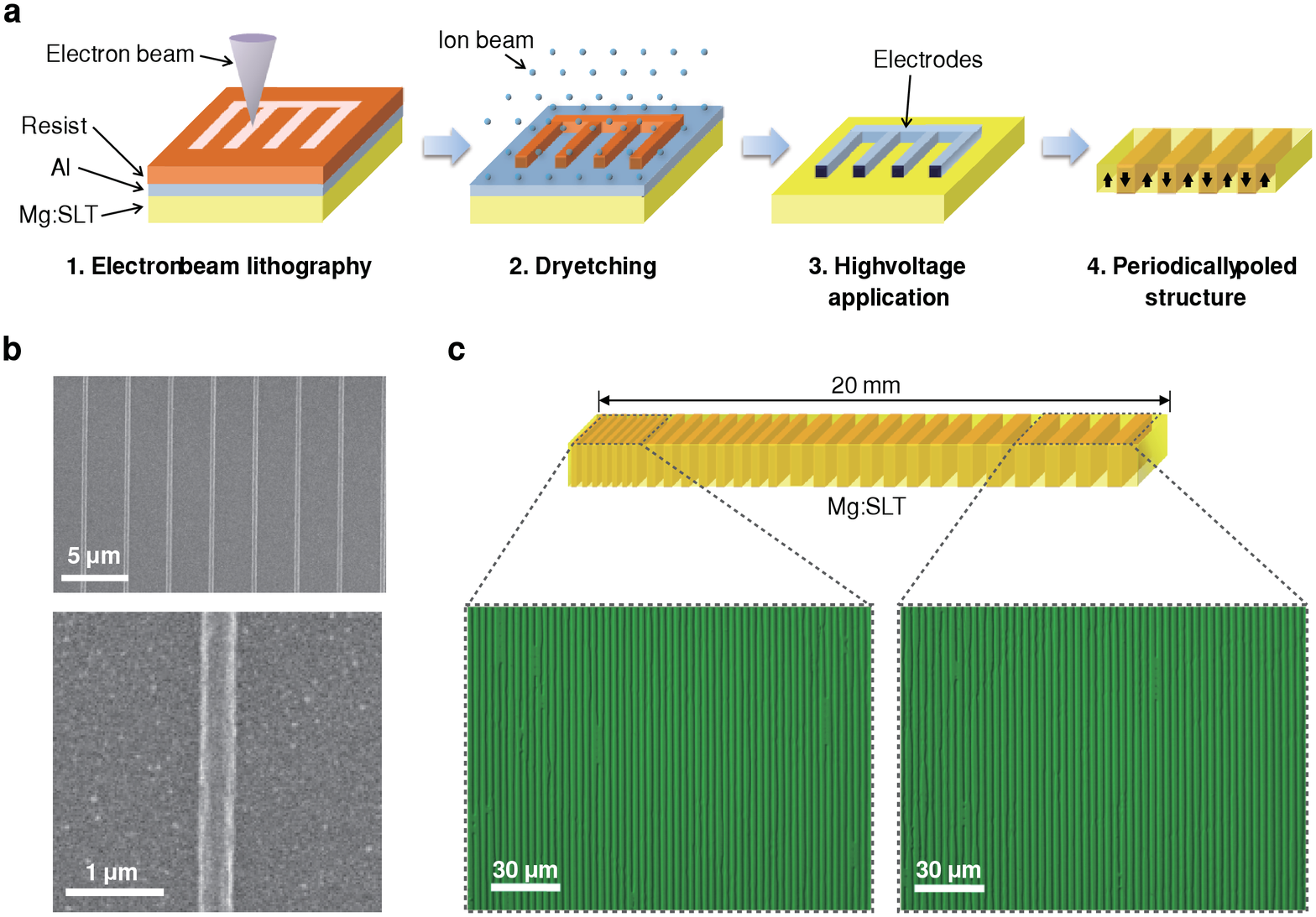}
\caption{Fabrication of chirped QPM lithium tantalate device 
using nano-electrode-poling technique. 
(a), Device fabrication process. 
Black arrows indicate the direction of ferroelectric spontaneous polarization 
in the periodically poled device.
Mg:SLT is Mg-doped stoichiometric lithium tantalate.
(b), Scanning electron microscopy images (upper and lower) 
of the 400 nm wide Al electrodes
fabricated for the poling period of 3.2 $\mu$m. 
(c), Optical Microscopy images of periodically poled structures. 
The QPM period varies from 3.12 $\mu$m (left) to 3.34 $\mu$m (right) 
along the 20 mm device length.}
\label{Fig2}
\end{center}
\end{figure}

The 1st-order QPM device was developed
using a $`$nano-electrode-poling$'$ technique.
Note that the conversion efficiency of the 1st-order QPM \cite{Fejer1992} is
supposed to be almost one order magnitude (9 times) larger than that of the 3rd-order QPM
\cite{{Nasr2008},{Nasr2008-2}};
however, it is impossible to realize
using conventional photolithography techniques.
For high-power ultraviolet pumping, 
Mg-doped stoichiometric lithium tantalate (Mg:SLT) 
\cite{{Yu2004},{Lim2013}} was selected
because it has a short absorption edge around 270 nm \cite{Nalwa2001}
and high thermal conductivity \cite{{Tovstonog2008},{Lim2011}}.
Furthermore, Mg:SLT is free from visible-light-induced nonlinear infrared absorption,
which is observed in Mg:LiNbO$_3$ \cite{Hirohashi2008}.
The new device fabrication process that was employed is shown in Figure \ref{Fig2}a.
A 500 $\mu$m thick Mg:SLT wafer doped with 1.0 mol\% Mg was used 
to suppress photorefractive damage and 
a 100 nm thick Al film was evaporated onto the wafer surface.
A nanoscale resist pattern was defined using electron beam lithography (EBL)
and then transferred to an Al pattern by dry etching on the surface of the Mg:SLT wafer.
EBL free from the diffraction limit of light
provides over 10 times higher accuracy than photolithography,
but special care is required to avoid charge-up in the ferroelectric Mg:SLT substrate.
The width of the Al electrode was 400 nm and
the duty ratio of the period was approximately 10\% to the period
assuming reasonable sidewise expansion 
(scanning electron microscopy (SEM) images are shown in Figure \ref{Fig2}b).
An electric field of 2 kV/mm was then applied in the vacuum chamber
to achieve a high electric field contrast
by suppression of the surface screening charge
for precise control of the domain duty ratio.
Domain sidewise motion was successfully controlled over 6,000 domains 
along the 20 mm device length with a 6.7\% chirped poling period varying
from 3.12 to 3.34 $\mu$m 
(optical microscopy images are shown in Figure \ref{Fig2}c). 
Although each EBL scanning area is limited 
to a width of 200 $\mu$m ($\ll$ 20 mm device length),
the extremely accurate translation stage enables a 20 nm connection error,
which results in a low phase slip between scanned areas.
The fluctuation in the domain duty ratio (typically 0.65)
was less than 10\% to the period (Figure \ref{Fig2}c), 
which suggests uniform conversion efficiency along the entire spectral range.

\begin{figure}[ht]
\begin{center}
\includegraphics[width=17.8cm]{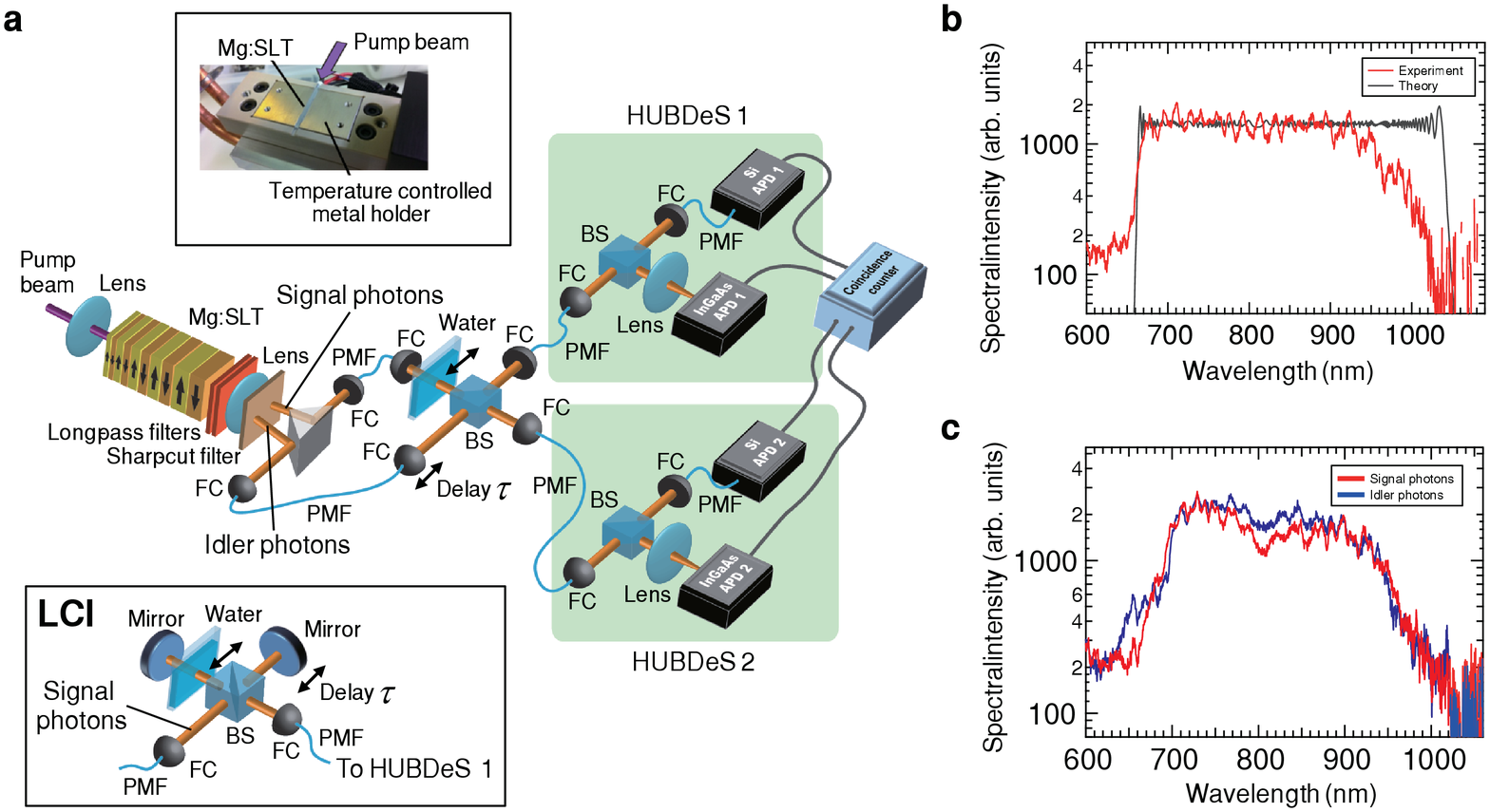}
\caption{Experimental setup. 
(a), TPI and LCI interferometers with hybrid ultra-broadband detection systems (HUBDeS). 
The upper inset shows the chirped QPM device 
set in the temperature controlled metal holder. 
HUBDeS consists of a Si avalanche photodiode (APD) and an InGaAs APD.
The lower inset shows the LCI interferometer. 
A 1 mm thickness of water can be inserted in the optical path 
in TPI and LCI interferometers.
BS, beam splitter; FC, fiber coupler; PMF, polarization-maintaining fiber;
Mg:SLT, Mg-doped stoichiometric lithium tantalate.
(b), Frequency spectrum of collinearly emitted photons from the device. 
Experimental data (red dots) and the theoretical curve (black line) are plotted. 
(c), Frequency spectra of photon pairs generated from the device in non-collinear emission. 
The observed data are plotted
for signal photons (red dots) and idler photons (blue dots). 
The transmission efficiency and detection efficiency of the spectrometer 
were calibrated (b,c.}
\label{Fig3}
\end{center}
\end{figure}

The experimental setup for the TPI is shown in Figure \ref{Fig3}a
(Details are given in Methods).
The chirped QPM Mg:SLT device is set in a temperature controlled metal holder
(Figure \ref{Fig3}a upper inset).
The temperature $T$ is controlled by a thermoelectric cooler with an accuracy of 0.1 K.
We have found that in ultra-high resolution regime
the linewidth of a pump laser degrades the dispersion tolerance \cite{Okano2013}. 
To ensure the dispersion tolerance,
we used a narrowband pump laser ($\sim$ 100 kHz)
with a wavelength of 401 nm.
The pump beam is focused to a diameter of 80 $\mu$m with a lens;
this diameter is sufficiently smaller than the cross-section of the device
(500 $\mu$m (vertical) $\times$ 800 $\mu$m (horizontal)),
and then cut by filters after the device.
Collinearly emitted photons had ultra-broad bandwidth
with a device temperature $T=351$ K
as shown in Figure \ref{Fig3}b.
The spectrum spans from 660 nm with a sharp rise.
We think that the offset in the shorter wavelength region ($\le$ 650 nm)
is stray fluorescence from the device.
Considering the frequency correlation 
with the center wavelength of 802 nm,
the spectrum should span up to 1040 nm with a bandwidth of 166 THz. 
The low intensity in the longer wavelength region ($\ge$ 950 nm)
may be due to the low coupling efficiency to the optical fiber 
for the spectrometer we used.
Figure \ref{Fig3}c shows non-collinearly emitted photon pairs 
(signal photons and idler photons)
with an emission angle of 1.5 degrees relative to the pump beam 
with a device temperature $T=353$ K.
With the detection event of a signal photon as a trigger, 
the coincidence count rate of the idler photons was 
typically 5\% (5$\times 10^5$ Hz) that of the single photon count rate
($\sim$ 1$\times 10^7$ Hz). 
Signal and idler photons coupled to fiber couplers (FCs)
are transferred to the TPI interferometer through polarization-maintaining fibers.
The delay $\tau$ is determined by the physical position of the FC. 
The coincidence count rate $C(\tau)$ is obtained by two HUBDeSs,
which consist of a Si avalanche photodiode (APD) with a detection bandwidth
spanning from 400 to 1060 nm
and an InGaAs APD with a bandwidth spanning from 950 to 1150 nm.
For the LCI experiments,
we used signal photons as a low-coherence light source 
with the exactly same bandwidth of the source for TPI
and the HUBDeS 1 for detection (Figure \ref{Fig3}a lower inset).
For the check of the dispersion effect,
a 1 mm thickness of water enclosed by two thin glass cover plates
is inserted in the optical paths of 
both TPI and LCI interferometers.
Note that according to the conventional definition for OCT \cite{Brezinski2006},
the interferograms below are expressed 
in units of the $`$delay$'$ $c\tau$/2, where $c$ is the speed of light,
considering the physical displacement of the delay mirrors
in OCT and QOCT systems (Figure \ref{Fig1}).

\begin{figure}[ht]
\begin{center}
\includegraphics[width=8.7cm]{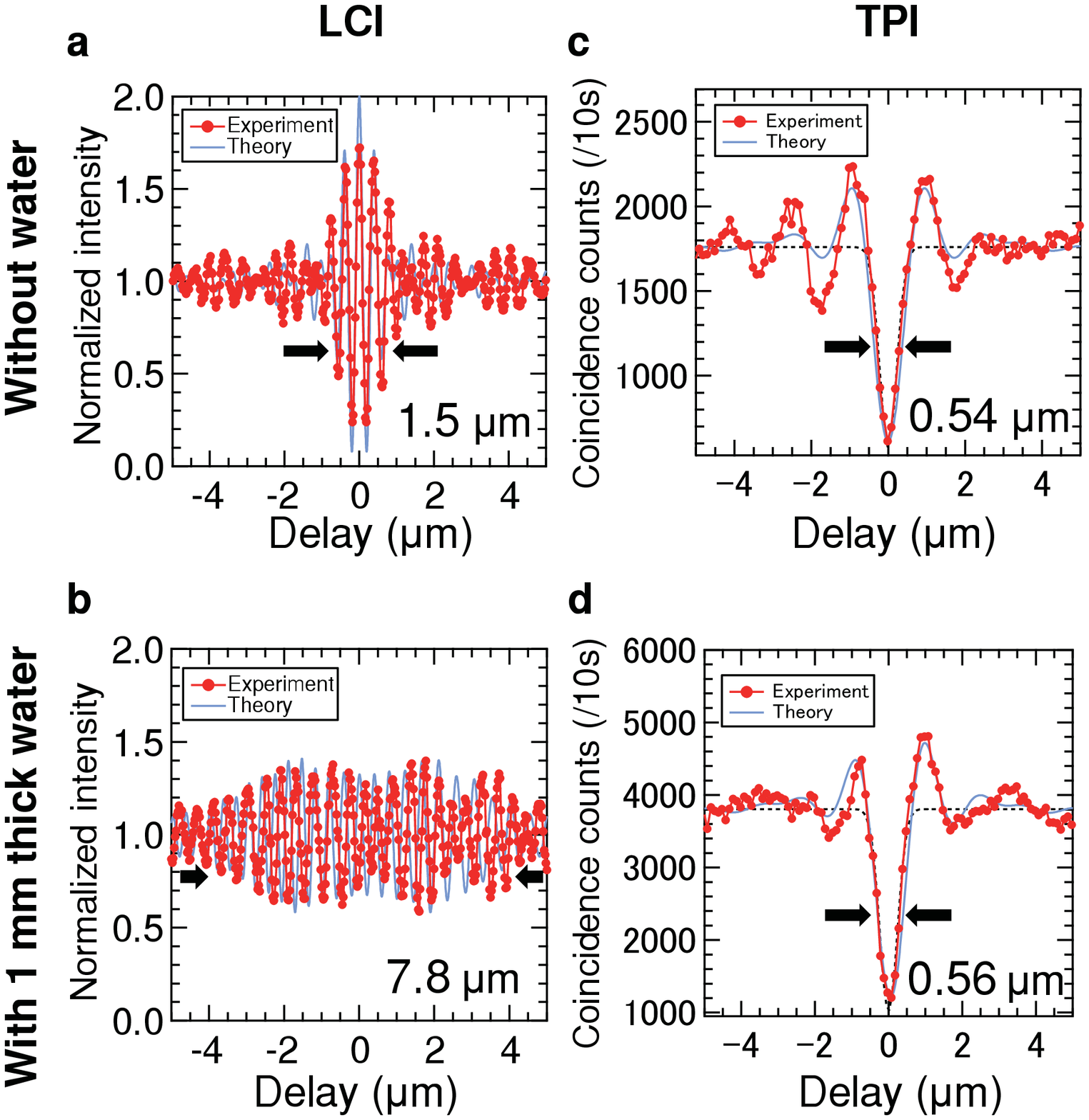}
\caption{Obtained LCI and TPI signals. 
(a,b), LCI fringes obtained in the LCI interferometer 
using signal photons as the source. 
The fringes without (a) and with (b) 
a 1 mm thickness of water inserted in the optical path. 
The experimental data (red dots) and the theoretical curves (blue line)
are plotted in units of the delay. 
The integration time was 1 second per point.
(c,d), TPI dips obtained in the TPI interferometer. 
The dips without (c) and with (d) the 1 mm thickness of water. 
The experimental data (red dots), the theoretical curves (blue lines)
and the Gaussian fitting curves (black dashed lines) 
are plotted in units of the delay. 
The integration time was 10 seconds per point. 
The red lines connecting the data points are a guide to the eye 
(a--d).}
\label{Fig4}
\end{center}
\end{figure}

The experimental results are shown in Figure \ref{Fig4}. 
First, the LCI signal was obtained as plotted by the red dots in Figure \ref{Fig4}a.
The full width at half maximum (FWHM) of the interference fringe is 1.5 $\mu$m,
which is slightly larger than the FWHM of 1.1 $\mu$m 
for theoretical calculation (blue line in Figure \ref{Fig4}a) 
assuming a rectangular spectral shape with a 166 THz bandwidth.
We think that this degradation is induced by the GVD:
When the thicknesses of the glass of BS in the optical paths
of the probe beam and the reference beam (Figure \ref{Fig3}a lower inset)
are different, the degradation occurs by the GVD of the glass.
We have found that the difference of the FWHMs can be explained
when we assume that the difference of the thicknesses of the glass is just 150 $\mu$m.

Figure \ref{Fig4}b shows the LCI signal
when a 1 mm thickness of water is inserted in the optical path.
Due to the GVD of the 1 mm thickness of water \cite{VanEngen1998},
the interference fringe becomes much broader.
The FWHM of the fringe is 7.8 $\mu$m,
which is more than 5 times larger than that without the 1 mm thickness of water sample. 
These results illustrate how GVD effect becomes crucial
in ultra-high resolution regime.

The main results of this paper are shown in Figures. \ref{Fig4}c and \ref{Fig4}d.
Figure \ref{Fig4}c shows the TPI signal without the water sample.
The FWHM of the observed HOM dip is 0.54$\pm$0.05 $\mu$m,
which surpasses 0.75 $\mu$m of LCI \cite{Povazay2002} for OCT 
using a broadband laser light with a center wavelength of 725 nm, 
which is the current best record resolution as far as we know.
The resolution of 0.54 $\mu$m in air
corresponds to the resolution of 0.40 $\mu$m
in water or biological tissue.
The theoretical curve (blue line in Figure \ref{Fig4}c)
assuming the observed spectrum of the signal photons (Figure \ref{Fig3}c)
fits well with the experimental data.
Note that for the same bandwidth of the photon source 
with a rectangular shaped spectrum,
the FWHM of the TPI is a half of the LCI \cite{Okano2013},
which is also an advantage of QOCT
especially when the bandwidth of the optical window is limited.

Finally, experimental demonstration of the dispersion cancellation 
of an ultra-high resolution TPI for QOCT is shown in Figure \ref{Fig4}d.
The HOM dip is almost unchanged from Figure \ref{Fig4}c
even when a 1 mm thickness of water is inserted in the optical path,
which is in striking contrast to the case of LCI (Figures. \ref{Fig4}a and \ref{Fig4}b).
The FWHM of the HOM dip is 0.56$\pm$0.04 $\mu$m
and the difference between the FWHMs with and without the 1 mm thickness of water
is 0.02 $\mu$m, which is smaller than the margins of errors.
The theoretical curve (blue line in Figure \ref{Fig4}d)
calculated taking the higher-order dispersion 
of the 1 mm thickness of water \cite{VanEngen1998} into account
fits also well with the experimental data.
Note that the asymmetricity of the HOM dip observed in the experimental data
is due to the higher odd-order dispersion \cite{{Okano2013},{Okamoto2006}},
which can be also observed in the theoretical curve.
Our theoretical calculation suggests that 
the resolution will be still kept 0.60 $\mu$m
even when a thickness of water is increased to 3 mm.
For thicker medium,
$`$phantom$'$ can also be used to compensate the dispersion effect for QOCT,
similar to OCT.
In case of QOCT, the thickness of the phantom does not have to be
exactly the same with the medium due to the inherit dispersion cancellation
by frequency entanglement demonstrated here.
In case of OCT, on the other hand,
the thickness or the GVD of the phantom has to be exactly the same with the medium
as we discussed on Figure \ref{Fig4}a, which is technically very difficult.
The detail of the theoretical estimation of FWHMs and the calculation
are given in Methods. 

\section*{Conclusion}

In conclusion, 
we have achieved 0.54 $\mu$m resolution TPI with dispersion cancellation
for ultra-high resolution QOCT,
which surpasses the current record resolution 0.75 $\mu$m of LCI for OCT.
We developed a 1st-order chirped QPM Mg:SLT device
using a $`$nano-electrode-poling$'$ technique with EBL
for ultra-broadband entangled photon pair generation
(166 THz, $\lambda$ = 660-1040 nm).
A 6.7\% chirped poling period varying from 3.12 to 3.34 $\mu$m 
was fabricated in over 6,000 domains along a 20 mm long device.   
We constructed stable interferometers with HUBDeSs that consist of
commercially available single photon detectors operated at room temperature.
In addition,
dispersion cancellation was demonstrated in TPI
against a 1 mm thickness of water inserted in the optical path.
Almost no degradation in resolution was observed in TPI 
(from 0.54 $\mu$m to 0.56 $\mu$m),
whereas the LCI resolution was significantly degraded
(from 1.5 $\mu$m to 7.8 $\mu$m). 
These results will open the door 
to ultra-high resolution QOCT imaging with depth resolution
less than half a micrometer.
Such ultra-high resolution QOCT will be beneficial for many different areas,
for example, 
following up the change of the retinal thickness with the ultra-high resolution
will greatly help the early detection of glaucoma \cite{Asrani2003}.
For this end,
the increase in the flux of entangled photons is very important.
The flux of 0.3 $\mu$W has already been realized 
using bulk QPM devices \cite{{Dayan2005},{Sensarn2009},{Tanaka2012}}.
Further increase in flux up to tens of $\mu$W
can be expected using slab or ridge type waveguide structures
\cite{{Kurimura2006},{Kou2011}}.
A promising direction may be an OCT/QOCT hybrid system
where OCT is used for a quick wide-range scan
and QOCT is used when ultra-high resolution / high precision observation is required.
We also note that the ultra-high resolution TPI with dispersion cancellation
demonstrated here is useful for quantum protocols,
including not only QOCT
but also quantum clock synchronization \cite{Giovannetti2001}, 
time-frequency entanglement measurement \cite{Hofmann2013}
and multimode frequency entanglement \cite{Mikhailova2008}.

\section*{Methods}

\subsection*{Details of experimental setup}
The pump laser system consists of 
a single-frequency continuous wave Ti:sapphire laser (MBR-110, Coherent) 
excited by a diode-pumped solid state laser (Verdi G-10, Coherent) 
and a resonant frequency doubling unit (MBD-200, Coherent). 
The output of the Ti:sapphire laser 
(wavelength: 802 nm; linewidth: approximately 100 kHz) 
is frequency doubled by second-harmonic generation. 
It is used as the pump beam with a power of 100 mW. 
This narrow linewidth ($\sim$ 100 kHz)
ensures the dispersion tolerance of TPI
in ultra-high resolution regime \cite{Okano2013}.
We treated the narrowband pump beam
as a monochromatic pump for numerical calculations 
based on the theory \cite{{Okano2013},{Okamoto2006}}. 
The focused pump beam at the chirped quasi-phase-matched device
has a confocal parameter (i.e., twice the Rayleigh length) of 18 mm. 

For the measurements of frequency spectra,
generated entangled photons from the device
are coupled to the the polarization-maintaining fiber 
and sent to a 300-mm spectrometer 
with a 150-grooves/mm grating blazed at 800 nm 
(SP-2358, Princeton Instruments) and 
a charge coupled device (CCD) camera (Pixis:100BRX, Princeton Instruments). 
The transmission efficiency of the spectrometer
and the quantum efficiency of the CCD camera were calibrated
in Figures \ref{Fig3}b and \ref{Fig3}c.

Each of HUBDeS shown in Figure \ref{Fig3}a
consists of a Si APD (SPCM-AQRH-14, Perkin Elmer) 
and an InGaAs APD (id401, idQuantique). 
The coincidence count rate $C$ between two HUBDeSs 
is obtained by the sum of coincidence counts between two APDs
as $C=C_{\textrm{Si-1/Si-2}}+C_{\textrm{Si-1/InGaAs-2}}
+C_{\textrm{Si-2/InGaAs-1}}+C_{\textrm{InGaAs-1/InGaAs-2}}$,
where for example $C_{\textrm{Si-1/InGaAs-2}}$ is 
the coincidence count rate between Si APD 1 in HUBDeS 1 
and InGaAs APD 2 in HUBDeS 2.
The coincidence time window of the coincidence counter
(id800, idQuantique) was 2 ns.

\subsection*{Analysis of interference signals}
For LCI experimental data
(plotted by red dots in Figures \ref{Fig4}a and \ref{Fig4}b),
a FWHM of interference fringe
is determined by the full width
at the middle between a base line and a peak height.
The base line is the average of the whole interference fringe along the delay.
The peak height is calculated using the smallest or largest data point.
In theoretical calculations for LCI
(blue lines in Figures \ref{Fig4}a and \ref{Fig4}b),
we calculated based on the theory \cite{Okano2013}
assuming a rectangular spectral shape
with a center wavelength of 802 nm and a bandwidth of 166 THz for the light source.
In the case with a 1 mm thickness of water inserted in the optical path,
we assumed the 2nd-order dispersion and 
the 3rd-order dispersion of the water \cite{VanEngen1998}.
For TPI experimental data (red dots in Figures \ref{Fig4}c and \ref{Fig4}d),
a FWHM of interference dip is determined by the FWHM of a Gaussian fitting curve
(black dashed lines in Figures \ref{Fig4}c and \ref{Fig4}d).
The Gaussian fit uses the whole range of the dip along the delay
and assumes the experimental visibility. 
The experimental visibilities of dips with and without the water dispersion
are 0.67$\pm$0.05 and 0.73$\pm$0.04, respectively.
In theoretical calculations for TPI
(blue lines in Figures \ref{Fig4}c and \ref{Fig4}d),
we assumed the observed spectrum of the signal photons (Figure \ref{Fig3}c), 
experimental coincidence count rate and experimental visibilities.
We also assumed the GVD and the 3rd-order dispersion of the water \cite{VanEngen1998}
for the case with the 1 mm thickness of water inserted.


\section*{Acknowledgements}

The authors thank Shoichi Sakakihara (ISIR, Osaka Univ.) for water sample preparation, Toru Hirohata and Masamichi Yamanishi (Hamamatsu Photonics) for lending ultra-broadband photo multiplier tubes for us, Yu Eto and Akira Tanaka (Osaka Univ.) for helpful discussions. This work was supported in part by the Japan Science and Technology Agency CREST project (JST-CREST), the Quantum Cybernetics project of the Japan Society for the Promotion of Science (JSPS), the Funding Program for World-Leading Innovative R\&D on Science and Technology Program of JSPS (FIRST), Grant in-Aids from JSPS; the Project for Developing Innovation Systems of the Ministry of Education, Culture, Sports, Science, and Technology (MEXT), the Network Joint Research Center for Advanced Materials and Devices, the Research Foundation for Opto-Science and Technology, and the Global COE programme.

\section*{Author contributions statement}
Experiments, measurements and data analysis were performed by M.O. with assistance of R.O. and S.T. Device fabrication was carried out by H.H.L. and S.K. All authors discussed the results and commented on the manuscript at all stages. M.O., S.K. and S.T. wrote the manuscript. The project was supervised by N.N., S.K. and S.T.

\section*{Additional information}

\subsection*{Competing financial interests}
The authors declare no competing financial interests.

\end{document}